\documentclass[pra,amssymb,amsmath]{revtex4}

\usepackage{color}

\newcommand{\be}{\begin{equation}}
\newcommand{\ee}{\end{equation}}

\begin{document}

\title{Towards a Bell-Kochen-Specker theorem of identity}

\author{R.            Srikanth}          \email{srik@poornaprajna.org}
\affiliation{Poornaprajna Institute of Scientific Research, Bengaluru,
  India}  \affiliation{Raman  Research  Institute, Bengaluru,  India.}
\author{Debajyoti  Gangopadhyay}  \affiliation{Annada College,  Vinoba
  Bhave University, Hazaribag, Jharkhand, India}

\begin{abstract}
In  contrast   to  conventional,  dynamical   entanglement,  in  which
particles  with  definite   identity  have  uncertain  properties,  in
so-called    statistical    entanglement,    which   arises    between
indistinguishable  particles because of  quantum symmetry  rules, even
particle  identities  are  uncertain.   The  Bell  and  Kochen-Specker
theorems imply that quantum  properties either lack realism or possess
it with  a caveat of contextuality  or nonlocality.  In  the matter of
identity  of  multi-particle  states of  indistinguishable  particles,
these contrasting ontological attitudes  are mirrored by the ``bundle"
vs.  haecceity views.  We offer  some arguments aimed at importing the
above theorems  to the issue of  identity in quantum  theory, with the
conclusion   (under   certain   assumptions)  that   indistinguishable
particles  either lack  individualism or  possess a  definite identity
with a caveat of contextuality or nonlocality.
\end{abstract}

\maketitle

\section{Introduction}

In  classical  physics,  the   notion  of  particle  identity  is  not
problematic.   Particles are distinguishable  individuals that  can be
specified   via   the   Leibnizian   ``Principle   of   Indentity   of
indiscernibles" (PII).   Any particle  is ``one of  its kind"  (can be
defined in 2nd order  logic, via quantification over properties). They
possess haecceity  (primitive thisness) and can be  labelled, e.g., 1,
2, $\cdots$.

In  quantum  mechanics,  the  formalism demands  that  multi-partitite
states  exist  in which  any  two  or  more particles  are  physically
indistinguishable \textit{in principle}, implying a failure of PII.  A
system of indistinguishable bosons (resp., fermions) must be symmetric
(resp.,  anti-symmetric) under pairwise  exchange of  particle labels,
making these states entangled simply  by virtue of the symmetry rules,
rather  than any interaction,  as it  is with  conventional, dynamical
entanglement.  A subtlety here is that indistinguishable particles may
have a  definite relation among themselves: E.g.,  the two-boson state
$a^\dag_\uparrow a^\dag_\downarrow|{\rm  vac}\rangle$, rendered in the
first  quantization  langauge as  the  statistical entanglement  state
$\frac{1}{\sqrt{2}}(|{\uparrow}\rangle|{\downarrow}\rangle            +
|{\downarrow}\rangle|{\uparrow}\rangle)$, corresponds to the statement
``there  exist two  indistinguishable  particles of  \textit{opposite}
spin".  One  might hope that this  could be used to  tell one particle
from  the  other.   Unfortunately  this  cannot be  done  because  the
relational property of  oppositeness, corresponding to the proposition
``has spin opposite to the other",  in the absence of definite spin of
the ``other", does not lead to a monadic property that can distinguish
the two particles.

In conclusion, one is led either to non-individualism, i.e., particles
lacking  definite individual  identity, or  to  an \textit{unphysical}
haecceity.   This  situation paralells  somewhat  the implications  of
quantum  mechanics  for  models  purporting to  explain  the  apparent
randomness  of outcomes  of measurements  corresponding to  a physical
property,  such as  position or  spin  along a  given direction.   The
difference  is   that  the  uncertainty  here   pertains  to  particle
\textit{identity} rather  than to a \textit{property}  that a particle
of given identity possesses.

As applied to possessed properties, this uncertainty was considered by
Einstein,  Podolsky   and  Rosen  \cite{epr}  as   indicative  of  the
incompleteness of quantum mechanics  (QM). They expressed hope for the
possibility   of   completing  QM   with   variables  that   possessed
\textit{realism},  i.e.,  definite   values  of  properties  prior  to
measurement, which  are revealed, and not  created during measurement.
The  definitive answer to  this problem  in standard  QM are  the Bell
\cite{bell,chsh,newbell}  and  Kochen-Specker  (KS)  \cite{ks,cabello}
theorems, which assert that  in certain situations, quantum mechanical
properties  either lack  realism or  are realistic  (governed  by some
hidden  variables  theory)   with  qualifications  of  nonlocality  or
contextuality, respectively.  We will generally refer to the import of
these theorems  as the uncertainty of possessed  properties of quantum
systems.   For reasons  we clarify  elsewhere, both  theorems  will be
regarded here  as facets  of the same  quantum phenomenon.  It  may be
said that this uncertainty of  identity, which has recently received a
lot  of attention  \cite{sancho,krause,dieks,hol,conf},  heightens the
quantum `mystery' already evident  in the uncertainty of the possessed
property.

\section{Contextual realism of the possessed property\label{sec:prop}}

The contradiction obtained  by the Bell and KS  theorem has its origin
in  the fact  that  observables corresponding  to  properties are  not
required  to commute  in  QM,  rendering it  impossible  to embed  the
algebra  of  these observables  in  a  commutative  algebra, taken  to
represent the  classical structure  of the putative  hidden variables.
An illustrative example is discussed below.

The GHZ nonlocality proof \cite{ghz} considers a three-qubit system in
the state 
\begin{equation}
|\Psi\rangle =  \frac{1}{\sqrt{2}} (|\uparrow\uparrow\uparrow\rangle -
|\downarrow\downarrow\downarrow\rangle),
\label{eq:ghz}
\end{equation}
on  which the  mutually  commuting three-qubit  measurements $XYY$  or
$YXY$ or $YYX$  or $XXX$ may be performed, where $X$  and $Y$ refer to
the Pauli  operators along those  directions, and a tensor  product is
assumed  between any  two operators.   The state  $|\Psi\rangle$  is a
+1-eigenstate of  the first three operators, and  a $-1$-eigenstate of
the last.   A realist model  to explain quantum indeterminacy  of this
system requires  that there exist  definite values of the  $X_j$'s and
$Y_k$'s,  where  $j,k  =  x,y,z$,  indicate the  position  index,  and
particles  $a,  b$   and  $c$  are  localized  at   $x,  y$  and  $z$,
respectively.

That no such assignment is possible is seen from the following table:
\begin{equation}
\begin{array}{cccc}
X_x & Y_y & Y_z & \rightarrow +1 \\
Y_x & X_y & Y_z & \rightarrow +1 \\
Y_x & Y_y & X_z & \rightarrow +1 \\
X_x & X_y & X_z & \rightarrow -1
\end{array} 
\label{tab:ghz}
\end{equation}
Any  realist assignment of  $\pm 1$  to the  $X_j$'s and  $Y_k$'s will
yield a product of +1 along  the first three columns because there are
two  copies of  $X$ or  $Y$ along  each column.  The product  of these
products is $+1$, which contradicts the $-1$ obtained above.

However a nonlocal-realistic explanation is possible: for example, let
all $X$'s and $Y$'s  = 1, except $X_x$, which will be  +1 in first row
and $-1$ in the last.

\section{Alternate view: Contextual haecceity} 

We  wish to  use a  Bell-Kochen-Specker-like argument  to  clarify the
sense  in which  particles have  uncertain identities.   We  require a
multi-partite system, rather than a single-particle quantum system for
which the Kochen-Specker theorem is  usually proved.  It turns out the
that the  GHZ proof of  nonlocality \cite{ghz} for 3  particles, suits
our purpose.   One difference worth  noting is that properties  have a
corresponding   observable   that  can   make   them  definite   under
measurement,  whereas  no  such  \textit{identity  observable}  exists
within  the `language' of  quantum mechanics  \cite{identity}. Another
point to note is that  multi-partite interference cannot be thought of
as interference of individual particles caused by indistinguishability
\cite{pitt}.   Instead,  they  are  interferences of  two-particle  or
multi-particle   amplitudes.    For   example,   in   the   well-known
Hong-Ou-Mandel  interferometer, the  interference happens  between the
`both-transmit' and `both-reflect' amplitudes \cite{hom}.

The state  $|\Psi\rangle$ in Eq. (\ref{eq:ghz}) can  be interpreted as
the following 3-boson state in a quantum optical situation:
\begin{eqnarray}
|\Psi\rangle    &\equiv&    (a^\dag_{0,z}a^\dag_{0,y}a^\dag_{0,z}    -
a_{1,z}^\dag       a^\dag_{1,y}      a^\dag_{1,x})|\textrm{vac}\rangle
\nonumber\\ &=  & \frac{1}{\sqrt{2}} (|\uparrow\uparrow\uparrow\rangle
- |\downarrow\downarrow\downarrow\rangle)_{abc}                 \otimes
\frac{1}{\sqrt{6}}\Pi_{[x,y,z]}
|\psi_x\rangle|\psi_y\rangle|\psi_z\rangle,
\label{eq:identity}
\end{eqnarray}
which represents bosonic particles created at positions $x, y$ and $z$
in the  GHZ state, and $\Pi$  represents a permutation  of the spatial
wave  functions over all  formal particle  labels.  The  state remains
invariant   under   label   exchanges   $a  \leftrightarrow   b$,   $a
\leftrightarrow c$, $b \leftrightarrow c$.  Any spin measurement $O_x$
on the first  position should be represented as  a symmetrization over
all three formal particle labels as $O_x \otimes I_{b} \otimes I_{c} +
I_{a} \otimes  O_x \otimes I_{c}  + I_{a} \otimes I_{b}  \otimes O_x$.
Just  as quantum  superposition expresses  that a  particle's property
lacks definiteness, a state like Eq. (\ref{eq:identity}) says that the
in the 3-particle state,  single particle identities are not definite,
in fact, they are \textit{maximally} indefinite.
 
Suppose the 3 particles  have definite individual identity, bound with
a  definite  spin  value  but  there  is  freedom  to  nonlocally  (or
contextually) influence the particle  identity associated with a given
position.   That the  data  in Table  \ref{tab:ghz}  can be  explained
follows from the assignment table
\begin{equation}
\left[\begin{array}{ccc}
X_x & Y_y & Y_z \\
Y_x & X_y & Y_z \\
Y_x & Y_y & X_z  \\
X_x & X_y & X_z 
\end{array} 
\right] \leftarrow
\left[\begin{array}{ccc}
b & a & c \\
a & b & c \\
a & b & c \\
a & b & c 
\end{array}\right],
\label{eq:newghz}
\end{equation}
with spin value assignments being $X_b= Y_a = Y_b = -1$ and $X_a = X_c
= Y_c = +1$.  This says that the identity of the  particles at $x$ and
$y$ are $b$  and $a$, respectively, in the  situation corresponding to
the top  row, but the opposite  in the situation  corresponding to the
last row.

\section{Conclusion and Discussion}

The uncertainty of individual identity in quantum mechanics because of
the  symmetry rules,  among one  of the  earliest  known non-classical
features, is seen in this light to make quantum mechanics even weirder
in  a  world  already  familiar  with  entanglement.   While  identity
uncertainty is, in this sense, similar to uncertainty of the possessed
property, the former comes with features absent in the latter, and the
implications of  these would have to  studied.  As noted,  there is no
identity observable, let  alone its maximally non-commuting conjugate,
so  that  the role  of  non-commutativity,  central  to the  Bell  and
Kochen-Specker theorems,  is borrowed from the property  algebra. As a
result,  the contradiction  obtained here  requires an  identity being
bound  to a  definite value  of the  property.  If  this  binding were
relaxed, and  particle identities bound  with the positions,  then the
standard nonlocal  realistic explanation of the  GHZ contradiction, as
given  in Section  \ref{sec:prop},  is also  applicable.  Of  interest
would be a contradiction, if  it exists, between quantum mechanics and
a  local/non-contextual  assignment  of  identity, even  allowing  for
nonlocal/contextual  properties, perhaps  using  HOM interoferometers,
quantum erasers, etc.

\acknowledgments

We are thankful to Ms. Akshata H. Shenoy for helpful discussions.

\end{document}